\def\beq{\begin{equation}}
\def\eeq{\end{equation}}
\documentclass{article}
\usepackage{amsfonts}
\usepackage{amssymb}
\def\p{{\sf p}}

\def\x{{\sf x}}
\def\t{{\sf t}}

\def\y{{\sf y}}

\def\a{{\sf a}}

\def\X{{\sf X}}

\begin{document}
\title{Stability of Covariant Relativistic Quantum Theory}
\author{V. Wessels, W. N. Polyzou\footnote{This work supported 
in part by the U.S. Department of Energy, under contract DE-FG02-86ER40286}\\
}

\maketitle
\begin{abstract}

In this paper we study the relativistic quantum mechanical
interpretation of the solution of the inhomogeneous Euclidean
Bethe-Salpeter equation.  Our goal is to determine conditions on the
input to the Euclidean Bethe-Salpeter equation so the solution
can be used to construct a model Hilbert space and a dynamical
unitary representation of the Poincar\'e group.  We prove three
theorems that relate the stability of this construction to properties
of the kernel and driving term of the Bethe-Salpeter equation.  
The most interesting result is that the positivity of
the Hilbert space norm in the non-interacting theory is not stable
with respect to Euclidean covariant perturbations defined by
Bethe-Salpeter kernels.  The long-term goal of this work is to
understand which model Euclidean Green functions preserve the
underlying relativistic quantum theory of the original field theory.
Understanding the constraints imposed on the Green functions by the
existence of an underlying relativistic quantum theory is an important
consideration for formulating field-theory motivated relativistic
quantum models.
				   
\end{abstract}
\bigskip
\bigskip

\section{Introduction}

The purpose of this paper is to investigate the conditions for
field-theory motivated calculations based on covariant Euclidean Green
functions to be interpreted as relativistic quantum mechanical
theories.  The problem is to use the Euclidean Green functions to
construct a model Hilbert space and the dynamical unitary
representation of the Poincar\'e group on this space \cite{wigner}.
There are two long-term goals of this research.  The first goal is to
understand which field-theory motivated calculations can be
interpreted as relativistic quantum theories.  The second goal is to
learn how to formulate phenomenological relativistic Euclidean quantum
models which have a clear connection to an underlying quantum field
theory.

This paper focuses on Euclidean rather than Minkowski Green functions
for two reasons.  First, Euclidean Green functions are used
extensively in applications, including lattice discretizations of
quantum field theories \cite{Montvay}, perturbative quantum field
theory, and in the Schwinger-Dyson equations \cite{roberts}.  The second
reason for considering Euclidean Green functions is that they are
directly related to the underlying relativistic quantum theory.

The relation of Euclidean Green functions to relativistic quantum
theory is discussed in the literature on axiomatic field theory.  A
readable summary of this relationship can be found in section 1.3 of
\cite{Montvay}.  More complete treatments appear in the original
literature \cite{osterwalder1},\cite{frohlich1},
\cite{osterwalder2}.  Reference \cite{osterwalder1} contains a 
clear statement of assumptions needed to reconstruct a relativistic 
quantum theory from a collection of Euclidean Green functions.
Reference \cite{frohlich1} contains a concise alternative treatment
in terms of Euclidean generating functionals.   

This paper investigates the quantum mechanical interpretation of the
inhomogeneous Euclidean Bethe-Salpeter equation.  This is the simplest
model where the relation to an underlying relativistic quantum theory
can be addressed.  In addition, it is an important tool for making
field-theory motivated models of two-body systems.  The input to the
Bethe-Salpeter equation consists of the Bethe-Salpeter kernel and the
two-point Euclidean Green function.  In applications this input is unknown
and has to be modeled.  A reasonable goal is to find conditions
on the model input that are sufficient to construct a relativistic
quantum mechanical two-body model.  Even this modest goal turns out to
be a non-trivial problem.  In this paper we prove three theorems that
address the relation of the phenomenological input to the existence of
an underlying relativistic quantum theory.

The first two theorems establish an interesting result, which is that
if the driving term of the inhomogeneous Bethe-Salpeter equation for
the Green functions is constructed from the {\it free} Euclidean Green
functions, and the four-point Green function is constructed by solving
the Bethe-Salpeter equation with a small Euclidean covariant kernel,
the positivity of the resulting Hilbert space norm is unstable with
respect to small variations in the kernel.  This was contrary to our
expectation that a sufficiently small Euclidean covariant kernel would
always preserve the underlying quantum mechanical interpretation of
the non-interacting system.  The third theorem shows that the
instability proof breaks down if the free two-point function is
replaced by a more realistic two-point function that has a
K\"all\'en-Lehmann weight with a continuous component to its mass
spectrum.

\section{Background}

The Euclidean formulation of quantum field theory plays a central role
in this paper.  This section contains a summary of the
relationship between Euclidean quantum field theory, relativistic
quantum mechanics, and Minkowski Green functions.  While all of this
material can be found in the literature, the focus in this section is
on aspects of these relationships which are needed to formulate
relativistic quantum mechanical models which are not local field theories.

There are three classes of generalized functions that are important in
this paper.  The first class are the Wightman functions, which are
vacuum expectation values of products of local field operators:
\beq
W_n (x_1, \cdots, x_n )
:=  \langle 0 \vert \phi_1(x_1) \cdots 
\phi_n(x_n)  \vert 0 \rangle . 
\label{eq:AA} 
\eeq
The second class of generalized functions are the Minkowski Green
functions, which are time-ordered vacuum expectation values of
products of local field operators
\beq
G_n (x_1 , \cdots , x_n) := \langle 0 \vert T(\phi_1(x_1) \cdots 
\phi_n(x_n) ) \vert 0 \rangle .
\label{eq:AB}
\eeq
The third class of generalized functions are the Schwinger functions,
or Euclidean Green functions, which will be defined later as analytic
continuations of the Minkowski Green functions or Wightman functions.

The transformation properties of the Wightman functions and Minkowski Green
functions are needed to construct unitary representations of the 
Poincar\'e group.  The Wightman and Minkowski Green functions are
Poincar\'e covariant.  For the Wightman function this is a consequence
of the transformation properties of the fields and the invariance of
the vacuum.  For the Minkowski Green functions locality is also needed
to ensure the Poincar\'e invariance of the time-ordered product of fields.

The symmetry group for the fields is inhomogeneous $SL(2,C)$.  The
group $SL(2,C)$ is the covering group for the Lorentz group.  The
$SL(2,C)$ matrices $\pm \Lambda$ are related to the Lorentz
transformation $\Lambda^{\mu}{}_{\nu}$ by 
\beq
\Lambda^{\mu}{}_{\nu} = {1 \over 2 } \mbox{Tr} \left (\Lambda \sigma_{\mu}
(\Lambda^{*})^t \sigma_{\nu} \right ),
\label{eq:AC}
\eeq
where $\sigma_{\mu} = (I, \vec{\sigma})$ and $\vec{\sigma}$ are the Pauli 
matrices.

Covariant fields are multicomponent operator densities
that transform as finite dimensional representations, $D(\Lambda
,\Lambda^*)$, of $SL(2,C)$:
\beq
U(\Lambda ,a) \phi (x) U^{\dagger} (\Lambda ,a) = 
\phi (\Lambda x+a) D(\Lambda , \Lambda^*).  
\label{eq:AD}
\eeq
where $\Lambda$ is an element of $SL(2,C)$  and 
$\Lambda x+a$ is a short-hand notation for 
$\Lambda^{\mu}{}_{\nu} x^{\nu} +
a^{\mu}$.
In $SL(2,C)$ the $\Lambda$ and $\Lambda^*$ define inequivalent
representations and both are needed to construct the irreducible
representations of the Lorentz group and the corresponding
representations of the four-dimensional orthogonal group.
  
The resulting covariance properties of the Wightman functions
and Minkowski Green functions are
\beq
W_n (x_1, \cdots, x_n ) =
W_n (\Lambda x_1+a, \cdots, \Lambda x_n +a  )
\prod_{i=1}^n D_i(\Lambda, \Lambda^*) 
\label{eq:AE}
\eeq
and
\beq
G_n (x_1, \cdots, x_n ) =
G_n (\Lambda x_1+a, \cdots, \Lambda x_n +a  )
\prod_{i=1}^n D_i(\Lambda, \Lambda^*),
\label{eq:AF}
\eeq
respectively. 

The Hilbert space and dynamical unitary representation of the
Poincar\'e group \cite{wigner} of the field theory is determined by the
Wightman functions.  This is illustrated with the following example of
a vector obtained by applying the product of smeared field
operators to the physical vacuum:
\beq
\vert \Psi \rangle := \phi (f) \phi (g) \vert 0 \rangle
\qquad \phi (f) :=  \int d^4 x \phi (x) f (x)  
\label{eq:AG}
\eeq
where there is an implied sum over the components of $f$ and $\phi$.
The Hilbert space scalar product of $\vert \Psi \rangle$ 
and $\vert \Psi \rangle'$ is related to a 
four-point Wightman function by 
\[
\langle \Psi \vert \Psi' \rangle :=
\]
\beq
\int dx dx' dy dy' g^*(x') f^*(y')  W(x',y';y,x) f'(y) g'(x)  
\label{eq:AH}
\eeq
where
\beq
W(x',y';y,x):= \langle 0 \vert 
\phi_2^{\dagger}(x')\phi_1^{\dagger}(y')
\phi_1 (y)\phi_2(x) \vert 0 \rangle .
\label{eq:AI}
\eeq
It is normally assumed that polynomials in the fields evaluated on
Schwartz test functions applied to the vacuum are a dense set of
vectors in the Hilbert space.  Schwartz functions \cite{gelfand}
\cite{simon} are
infinitely differentiable functions which decrease faster than any
inverse polynomial.  Continuous multilinear functionals on products of
Schwartz functions are tempered distributions.  All of the generalized
functions in this paper are assumed to be tempered distributions.
This is a mild assumption which ensures the analytic properties
that are traditionally assumed in quantum field theories.

The scalar product of vectors constructed by applying more complex
polynomials of smeared fields applied to the vacuum can also be
expressed in terms of Wightman functions.
 
The transformation properties of the state (\ref{eq:AG}) 
follow from the covariance (\ref{eq:AD}) of the fields
\beq
U(\Lambda ,a) \vert \Psi \rangle = \phi_1 (D_1(\Lambda,\Lambda^*) f') 
\phi_2 (D_2(\Lambda,\Lambda^*) g') \vert 0 \rangle  
\label{eq:AJ}
\eeq
where $f'(x) = f(\Lambda^{-1}(x-a))$ and $g'(x) = g(\Lambda^{-1}(x-a))$.

The Poincar\'e invariance of the Hilbert space scalar product    
\beq
\langle U(\Lambda ,a) \Psi \vert  
U(\Lambda ,a) \Psi' \rangle = \langle \Psi \vert  
\Psi' \rangle 
\label{eq:AK}
\eeq 
follows if the test functions have the transformation 
property  
\beq
U(\Lambda ,a)f(x) = D(\Lambda,\Lambda^*) f(\Lambda^{-1}(x-a)).
\label{eq:AL}
\eeq
This illustrates how the covariance of the Wightman functions 
leads to a unitary representation of the Poincar\'e group on the 
Hilbert Space defined by the Wightman functions.

The structure of the Hilbert space scalar product in (\ref{eq:AH}) is
not limited to relativistic quantum field theory; it is the {\it
general form} of the Hilbert-space scalar product in any relativistic
quantum theory where the Poincar\'e group is implemented by manifestly
covariant \cite{wp1} transformations of test functions like $f$ and
$g$.  This suggests the possibility of formulating ``approximations''
to the underlying field theory that remain relativistic quantum
theories.
 
The construction of the relativistic quantum dynamics defined by
generalizations of (\ref{eq:AH}),(\ref{eq:AJ}) and (\ref{eq:AL},) can
be compared to the familiar textbook construction of Fock space
Poincar\'e generators based on Noether's theorem.  The Noether's
theorem construction is limited to free fields.  The introduction of
an interaction leads to ``generators'' that are not densely 
defined operators on the Fock space.  They do not have self-adjoint 
extensions and attempts to regularize these ``generators'' so they become
self-adjoint operators invariably lead to violations of the
commutation relations.  The Wightman-function construction agrees with
the textbook construction for free fields, and it is not limited to
perturbation theory or Lagrangian field theory for interacting
systems.

While the Wightman functions determine the quantum interpretation of
the field theory, they are difficult to use in models because they do
not have inverses.  This is apparent from the structure of the 
Fourier transform of the two-point Wightman function of a scalar 
field theory, which is, up to a constant,  the product of a mass-shell delta
function and a positive energy Heaviside function.  In spite of their
role in defining the underlying quantum theory, Wightman functions 
have not been extensively used in practical
applications.  

The Minkowski Green functions have the advantage that as Green
functions, they are expected to have inverses,
which make it possible to derive useful equations, like the
Bethe-Salpeter equation.  The disadvantage of these Green functions is
that they are not directly related to the underlying quantum theory.
Their connection to a quantum theory is that for each time-ordering
the Minkowski Green function agrees with the Wightman function whose
fields have the same order as the Minkowski times. The quantum theory
can be recovered if there is enough analyticity to reconstruct the
Wightman functions from the limited information contained in the
Minkowski Green function.
 
In quantum field theories the Euclidean Green functions or Schwinger
functions are related to the Wightman functions and Minkowski Green
functions by analytic continuation.  Schwinger functions are important
because (1) they exist under mild conditions (2) they can be used to
directly construct the underlying relativistic quantum theory and (3)
they are expected to have inverses.  The original use of analytic
continuation of the Fourier transforms of the Minkowski Green
functions to imaginary energies is due to Dyson
\cite{dyson}, who used the analytic continuation as a tool to study
ultraviolet divergences in perturbation theory.

In 1958 Schwinger \cite{schwinger} used the spectral condition of the
physical intermediate states and Poincar\'e covariance to show that
the $n$-point Minkowski Green functions of a quantum field theory
can be analytically continued to imaginary times:
\beq
S_n (\x_1, \cdots ,\x_n) := \lim_{\phi:0\to \pi/2} =
G_n (\vec{x}_1, e^{-i \phi} t_1 , \cdots , \vec{x}_n, e^{-i \phi} t_n).
\label{eq:AN}
\eeq
In equation (\ref{eq:AN}) and the remainder of this paper
$x=(x^0,\vec{x})$ denotes 
a real Minkowski four
vector and $\x=(\x^0,\vec{\x})$ denotes a real Euclidean four vector.
The components of these vectors are related by
\beq
\x_k := (\x^0 , \vec{\x}_k) = (-ix^0_k ,\vec{x}_k ) \qquad
x_k := (x^0 , \vec{x}_k) = (i\x^0_k ,\vec{\x}_k ). 
\label{eq:AO}
\eeq

This analytic continuation (\ref{eq:AN}) defines the Schwinger
functions for non-coincident times, $t_i\not= t_j$.  The existence of
this analytic continuation is based on a multi-variable generalization
\cite{wightman} of the Payley-Wiener-type theorem that states
that 
\beq
\tilde{f} (t-t') := \int dE f(E) e^{-i E (t-t')}
\label{eq:AOO}
\eeq
can be analytically continued to the lower-half $(t-t')$-plane if
$f(E)$ is a tempered distribution with support for positive energies.
In the definition (\ref{eq:AN}) the order of the Euclidean times is the
same as the order of the Minkowski times in the Minkowski Green
functions.  The covariance properties of the Schwinger functions 
$S_n$ are not apparent
from the definition (\ref{eq:AN}) and will be discussed separately.

The Fourier representation of $G_n (x_1, \cdots ,x_n)$ can be used to
extend the analytic continuation in (\ref{eq:AN}) to complex
$(x_{i}-x_{i+1})$ with $-(x_i-x_{i+1})\in R^4 + i {\cal V}^+$, where
${\cal V}^+$ is the open future-pointing light cone.  On this domain
the real part of $i p_i \cdot (x_{i}-x_{i+1})$ is negative when $p_i$
is time-like with positive energy, which ensures the analyticity.  This
domain is called the tube in the literature \cite{wightman2}.  In
principle each time ordering of the Minkowski Green function has a
different analytic continuation.  The different analytic continuations
are defined on disjoint domains characterized by different Euclidean
time orderings.  Also, because for each time ordering the Minkowski
Green function is equal to the Wightman function with the fields
ordered in the same order as the times, each analytic continuation is
the analytic continuation of a different Wightman function.

Since the Wightman functions transform covariantly with respect to a
finite dimensional representation of the Lorentz group, (\ref{eq:AE}),
the complexification of the covariance condition can be used to extend
the domain of analyticity from the tube to the domain generated from
the tube by complex Lorentz transformations, called the extended tube
\cite{wightman2}. 

This extension is done using the complex Lorentz group.  Complex
Lorentz transformations are complex linear transformations that leave
the Minkowski line element invariant.  In the appendix it is shown
that the most general complex Lorentz transformation connected to 
the identity has the representation
\beq
\Lambda (A,B)^{\mu}{}_{\nu} = {1 \over 2} \mbox{Tr} (A \sigma_{\mu} 
B^t \sigma_{\nu}) = \Lambda^{\mu}{}_{\nu} (A,B)
\label{eq:AP}
\eeq
where $A$ and $B$ are independent $SL(2,C)$ transformations.  The
covering group of the complex Lorentz groups is $SL(2,C)\times
SL(2,C)$.  Real Lorentz transformations are obtained by taking $A =
B^* = \Lambda$.  The connection with Euclidean covariance follows
because four-dimensional complex orthogonal transformations have a
representation similar to (\ref{eq:AP}),
\beq
E^{\mu}{}_{\nu}(A,B){} = {1 \over 2} \mbox{Tr} (A \tilde{\sigma}_{\mu} 
B^t \tilde{\sigma}^{\dagger} _{\nu})
\label{eq:AQ}
\eeq
where $\tilde{\sigma}_{\mu} = (i \sigma_0, \vec{\sigma})$.  In the
Euclidean case real $O(4)$ transformations have $A,B \in SU(2)$.  It
follows from (\ref{eq:AP}) and (\ref{eq:AQ}) that the covering 
group of the complex orthogonal group in four dimensions is also 
$SL(2,C)\times SL(2,C)$.  Complex
covariance can be used to extend the analytic continuation to all
points $x_i'= \Lambda(A,B)^{-1}(x_i-a)$ that can be reached from
points in the tube by complex Lorentz transformations using
\[
W_n (x_1' , \cdots, x_n')  =
\]
\beq
W_n (x_1, \cdots, x_n  ) \prod_{i=1}^n
D_i(\Lambda(A,B)) 
\label{eq:AR}
\eeq
which is consistent with real Lorentz covariance when $A=\Lambda =
B^*$ and with real Euclidean covariance when $A,B \in SU(2)$.
Restricting to real Euclidean points in the extended tube and letting
$A,B\in SU(2)$ leads to real Euclidean invariance of each of the $n!$
analytic continuations.

In quantum field theory the $n!$ analytic continuations are defined on
common domains consisting of open sets of real Minkowski space-like
separated points called Jost points \cite{wightman2}.  If the Wightman
functions are constructed from local fields the analytic continuations
have to be identical on the Jost points because the order of the
fields is irrelevant (up to sign for fermions) on space-like separated
points.  The result is a that all of the $n!$ analytic continuations
are part of a single-valued Euclidean covariant Green function with
transformation properties
\[
S_n (\x_1' , \cdots, \x_n')  =
\]
\beq
S_n (\x_1, \cdots, \x_n  ) \prod_{i=1}^n
D_i(\Lambda(A,B)) 
\label{eq:ARR}
\eeq
with $\x^{\mu \prime } = E(A,B)^{\mu}{}_{\nu} \x^{\nu} + \a^{\mu}$. 
This defines what we mean by Euclidean covariance. 

The real Euclidean transformation properties of the Schwinger
functions are obtained from those of the corresponding Wightman
functions by replacing the pair $(\Lambda, \Lambda^*)$ by  
the $SU(2)$ matrices $(A,B)$:
\beq
D(\Lambda , \Lambda^*) \to D(A,B)
\eeq
An important property of the Schwinger functions is that they can be
used to directly formulate the underlying quantum theory.  This was
first done in 1973 by Nelson
\cite{nelson}, who also related the Schwinger functions to moments of
a Euclidean functional integral.  A more useful construction was given
during the same year by Osterwalder and 
Schrader\cite{osterwalder1}\cite{osterwalder2}.
Fr\"ohlich \cite{frohlich1} gave an elegant reformulation of 
the Osterwalder-Schrader construction in terms of Euclidean 
generating functionals.     
In \cite{osterwalder1} Osterwalder and Schrader
identified properties of Schwinger functions that are sufficient to
reconstruct the underlying relativistic quantum field theory.
They exploited the relation between the Schwinger 
functions and the Wightman functions,
which are the boundary values of the analytic continuation:
\beq
W_n (x_1 \cdots x_n ) = \lim_{\x_1^0 > \cdots > \x_n^0 \to 0}  
S_n (\vec{\x}_1, \x_1^0 - i x^0_1 , \cdots ,  
\vec{\x}_n, \x_n^0 - i x^0_n ).
\label{eq:AT}
\eeq
There are $n!$ Wightman functions depending on the ordering of the $n$
fields; the ordering on the Euclidean times in the limit (\ref{eq:AT})
selects the Wightman function with the fields ordered in the {\it
same} order as the Euclidean times.  In equation (\ref{eq:AT}) there are no
restrictions on the Minkowski times in the resulting Wightman
function.  The other $n!-1$ Wightman functions are selected by taking
the limit of the Schwinger function with different orderings on the
Euclidean times.  The $n!$ Wightman functions agree with 
$n!$ analytic functions generated by the $n!$ time orderings 
of the Minkowski Green functions.    

By exploiting this relationship Osterwalder and Schrader were able to
find conditions on the Schwinger functions that are, up to some
technical growth conditions \cite{osterwalder2}, equivalent to the
axioms given by Wightman.  Furthermore, Osterwalder and Schrader were
able to identify an independent subset of axioms \cite{osterwalder1}
that are sufficient to construct the Hilbert space and a unitary
representation of the Poincar\'e group with four momentum satisfying a
spectral condition.

The relevant conditions on the Schwinger functions are that they
should be Euclidean covariant (\ref{eq:ARR}) tempered distributions
which satisfy a property called reflection positivity.  Reflection 
positivity will be discussed more completely in the next section.   
It is used to construct the physical Hilbert space scalar product
and is most simply illustrated with an example that is analogous to 
the example (\ref{eq:AH}). 
In the Osterwalder Schrader approach the scalar 
product (\ref{eq:AH}) is replaced by   
\[
\langle \Psi \vert \Psi' \rangle =
\]
\beq
\int d\x d\x' d\y d\y' g^*(\theta \x') f^*(\theta \y')  
S(\x',\y';\y,\x) f'(\y) g'(\x)  
\label{eq:AV}
\eeq
where $f(\x),g(\x)$ are functions of four Euclidean space-time
variables with disjoint positive-Euclidean-time support for $0< \y^0 <
a < x^0 < b < \infty$ and $\theta \x= (-\x^0, \vec{\x})$ is the
Euclidean time-reversal operator.  Here $a$ and $b$ are constants that
serve to separate the Euclidean time support of $f$ and $g$.  This
sesquilinear form is the physical Hilbert space scalar product in the
Osterwalder Schrader formalism.  In field theories the ordering on the
support of the Euclidean times is equivalent to choosing an order of
the fields in the Wightman functions.  Reflection positivity requires
that (\ref{eq:AV}) is non-negative when $\vert \Psi \rangle = \vert
\Psi'
\rangle$.  Like the Wightman case there are
generalizations of (\ref{eq:AV}) to more complicated states.  The
surprising feature of (\ref{eq:AV}) is the {\it physical Hilbert space
can be directly defined in terms of the Schwinger functions, with no
need to transform back to the Minkowski formulation of the Hilbert
space based on Wightman functions}.

In the next section we discuss how the Euclidean covariance of
reflection positive Schwinger functions can be used to construct a
unitary representation of the Poincar\'e group on the Hilbert space
defined by the scalar product (\ref{eq:AV}).

This paper is concerned with a relativistic quantum theory which is
not necessarily a local quantum field theory.  In applications where
locality is relaxed, assumptions need to be made about the
relationship between the Wightman functions, the Minkowski Green
functions, and the Schwinger functions.  In the absence of locality
the $n!$ Wightman functions are no longer required to be related.  The
Minkowski Green function does not necessarily have a single valued
analytic continuation.  To proceed it is useful to assign a
fundamental significance to one of the three classes of generalized
functions and use this to formulate a quantum theory.

In this paper we always assume that the model Schwinger functions have
a more fundamental status.  The model Schwinger functions
are taken to be single valued tempered distributions which
satisfy the covariance condition (\ref{eq:ARR}).  This requirement is
motivated by the observation that most of the models of interest lead
to single-valued Euclidean covariant Schwinger functions.  For
example, models generated from Euclidean generating functionals and
approximations to Euclidean path integrals naturally lead to single
valued Euclidean covariant model Schwinger functions. 

Since the model Schwinger functions will not have all of the
properties of the Schwinger functions of a local field theory, the
relation to model Minkowski Green functions and model Wightman
functions cannot be expected to be identical to the relationship
found in local field theories.  The most important requirement is that
the model Schwinger functions will also be assumed to have enough
reflection positivity to build a relativistic quantum theory with the
desired particle content.  The precise formulation of this condition 
will be discussed in sections three and four.

The focus of this paper is properties of two- and four-point Euclidean
Green functions.  Reflection positivity imposes conditions on the two
and four-point function for them to be part of a system of Schwinger
functions that define the Hilbert-space scalar product of a
relativistic quantum theory.

The Bethe-Salpeter equation can be understood by considering the 
cluster property of the four-point Schwinger function, which  
has the form
\beq
S= S_0 + S_t
\label{eq:AAJ}
\eeq
where $S_0$ is a sum of products of two-point Schwinger functions and
$S_t$ is the truncated four-point Schwinger function.  If $S$ and
$S_0$ can also be understood as Green functions in the traditional
sense, i.e. as kernels of integral operators with inverses, then it is
possible to define the Bethe-Salpeter Kernel: 
\beq
K= S_0^{-1} S_t S^{-1}. 
\eeq
When $K \not=0$ the four-point Schwinger function $S$ is generated by
solving the inhomogeneous Euclidean Bethe-Salpeter equation
\beq
S = S_0 +  S_0 K S  
\label{eq:AAK}
\eeq
given a Bethe-Salpeter kernel $K$.  

In quantum field theory $S$, $K$, and $S_0$ are only known formally and
the Bethe-Salpeter equation is a constraint that relates these three
quantities.  In order to make this into a solvable equation two
modifications are normally made.  First, $S_0$ is replaced by the
$S_0$ of a free field theory. The virtue of this ``approximation'' is
that the resulting $S_0$ is known.  Second, the Bethe-Salpeter kernel
$K$ is modeled, using either perturbative methods or theoretically
and/or experimentally motivated phenomenological methods.  The 
cluster properties (\ref{eq:AAJ}) suggest that the kernel should be 
a short-ranged operator.
 
In this paper we examine the following stability question.  Let
$K$ be a sufficiently small, Euclidean covariant, model Bethe-Salpeter
kernel and let $S_0$ be the Schwinger function of an underlying
relativistic quantum theory.  Is the solution, $S$, of the Euclidean
Bethe-Salpeter equation (\ref{eq:AAK}) the $S$ of a relativistic
quantum theory?  In the absence of such a stability, arbitrarily small
uncertainties in the model Bethe-Salpeter Kernel could lead to a
theory that is no longer a relativistic quantum theory.  In this paper
we show that for the special case that $S_0$ is the Schwinger function
of a {\it free} field theory that $S$ can fail to satisfy the
constraints imposed by reflection positivity for arbitrarily small
Euclidean-covariant kernels $K$.  While this result does not apply to
the exact Bethe-Salpeter equation, where $S_0$ is {\it not} the free
$S_0$, many applications of the Bethe-Salpeter equation ``approximate''
$S_0$ by the free $S_0$.

In the next section we summarize the structure of a Euclidean
Relativistic quantum theory.  We define reflection positivity and
review how it is used to construct the physical Hilbert space of the
theory and the relativistic quantum dynamics.  We identify necessary
conditions for reflection positivity that we use to study four-point
Schwinger functions in section five.  The construction of a
free-particle dynamics is illustrated in section four.  In section
five we show that the necessary conditions for reflection positivity,
which are derived in section three, can be violated for arbitrarily
small Bethe-Salpeter kernels when $S_0$ is the $S_0$ of a free-field
theory. We also prove a result that suggests that the instability 
may not appear in the exact Bethe-Salpeter equation.  
The implications of this result are discussed in section
six.

\section{Euclidean Relativistic Quantum Mechanics}

In this section we give a short description of Euclidean relativistic
quantum theory.  We review how a collection of Euclidean-invariant
reflection-positive Schwinger functions are used to construct the
physical Hilbert space and the dynamical unitary representation of the
Poincar\'e group of a relativistic quantum theory.  A readable
description of the main elements of this construction can can be found
in \cite{Montvay}.  Mathematical treatments of this construction can
be found in
\cite{osterwalder1}\cite{osterwalder2}\cite{frohlich1}
\cite{frohlich2}.  Self-adjointness of the boost generators
can be established using the methods
discussed in \cite{klein1}\cite{klein2}\cite{frohlich3}. 

For the purpose of illustration we consider the case of Schwinger
functions for a scalar field.  We discuss the spin $1/2$ case
in section three.

A relativistic quantum theory is defined by a unitary representation
of the Poincar\'e group \cite{wigner} acting on the physical Hilbert
space with four-momentum generators that have a spectrum in the 
future-pointing light cone.

As discussed in section two the physical Hilbert space of Euclidean
relativistic quantum mechanics is defined by constructing an inner
product on a nice set of vectors; limits are used to complete the
Hilbert space.  As in the Wightman case, the Euclidean Green functions
are assumed to yield finite results when integrated against Schwartz
test functions \cite{gelfand} of $4N$ Euclidean space-time variables.

Let ${\cal S}$ be the space of finite sequences of Schwartz test functions
in different numbers of Euclidean space-time variables:
\beq
\langle \x \vert f \rangle := \{f_0,f_1 (\x_{11}), 
f_2 (\x_{21},\x_{22}), \cdots ,
f_k (\x_{k1},\cdots , \x_{kk}) \}
\label{eq:BA}
\eeq
where $f_l (\x_{l1},\cdots, \x_{ll})$ is a Schwartz function in $l$ Euclidean 
space-time variables.  These functions are the Euclidean replacements for
the functions $f$ and $g$ that appear in the Minkowski scalar product
(\ref{eq:AH}). 
 
To construct the physical Hilbert space Osterwalder and Schrader
\cite{osterwalder1}
introduce the subspace ${\cal S}_>$ of ${\cal S}$, where each of the
functions $f_l$ has support for an ordered set of positive Euclidean
times, $\x^0_{ll} > \cdots > \x^0_{l1} >0 $ .  The projection on
${\cal S}_>$ is denoted by $\Pi_>$.  The space ${\cal S}_>$ is natural
for two reasons.  First, each Euclidean time-ordering defines a 
scalar product that is equal to the corresponding Minkowski scalar product 
defined in terms of the Wightman function with the fields
ordered in the same order as the Euclidean times.  Second, the ordering
has a well-defined Minkowski limit if the Wightman functions
are consistent with requirements imposed by the spectral condition. 

The Euclidean time-reversal operator $\Theta$ on ${\cal S}$ is 
defined by 
\beq
\langle \x \vert \Theta f \rangle  := \{f_0,f_1 (\theta \x_{11}), 
f_2 (\theta \x_{21},\theta \x_{22}), \cdots , 
f_k (\theta \x_{k1},\cdots , \theta \x_{kk})\}.
\label{eq:BB}
\eeq
where $\theta (\x^0, \vec{\x}):= (-\x^0, \vec{\x})$.   

Given a collection of Euclidean-covariant Schwinger functions
\beq
\{ S_n (\x_1, \cdots, \x_n ) \} ,
\label{eq:BC}
\eeq
which are tempered distributions,  and test functions 
$f, g \in {\cal S}_>$, Osterwalder and Schrader define the quadratic form
\[
(\Theta f, S g) =  (f, \Theta S g) := 
\]
\beq
\sum_{m,n} \int d^4\x_1 \cdots d^4\x_{m+n} 
f^*_m (\theta \x_m, \cdots, \theta \x_1 ) 
S_{m+n} (\x_1, \cdots ,\x_{m+n} )
g_n (\x_{m+1}, \cdots , \x_{m+n} ).
\label{eq:BD}
\eeq
The support conditions on the functions $f$ and $g$ select 
the part of the Schwinger function that has a given Wightman 
function as the boundary value of an analytic function.

The relation $f \sim g$ if and only if 
\beq
\left (\Theta (f-g), S( f-g)\right ) = 0 .
\label{eq:BE}
\eeq
defines an equivalence relation on ${\cal S}_>$. 
The functions $f\in {\cal S}_>$ can be put into disjoint 
equivalence classes with respect to this equivalence relation;
the equivalence class containing 
$f \in {\cal S}_>$ is denoted by $[f]_\sim$.  The equivalence class containing 
zero is denoted by $[0]_\sim$.  The equivalence relation is $S$ dependent.

The sesquilinear form (\ref{eq:BD}) is well-defined on equivalence classes: 
\beq
\langle [f]_{\sim} \vert [g]_\sim \rangle = 
(\Theta f ,S g) 
\label{eq:BF}
\eeq
where $f$ and $g$ are any representatives of $[f]_\sim$ and $[g]_\sim$ 
respectively.

{\it Reflection positivity} is the condition that 
\beq
\Vert [f]_\sim \Vert^2 := \langle [f]_{\sim} \vert [f]_\sim \rangle = 
(\Theta f ,S f) \geq 0
\label{eq:BG}
\eeq
and vanishes only for $[f]_\sim= [0]_\sim$. 

A dense set of vectors in the physical Hilbert space is the space of
equivalence classes of functions $[f]_\sim \in {\cal S}_>$ .  The physical
Hilbert-space inner product of two vectors is given by (\ref{eq:BF}),
where the inner product can be evaluated using any $f
\in [f]_\sim$ and $g \in [g]_\sim$.  The physical Hilbert space is
obtained by completing the space of equivalence classes in the norm
defined by (\ref{eq:BG}).

This defines the physical Hilbert space directly in terms of the Schwinger
functions.  {\it Reflection positivity is equivalent to the statement that
vectors in the physical Hilbert space have positive length}. 

The involution, $\Theta$, on the Euclidean space serves as a ``conjugation 
operator''.  We will show how this ``conjugation'' converts a 
representation of a subgroup
of the complex Euclidean group into a unitary representation of the 
Poincar\'e group. 

A {\it necessary} condition for reflection positivity is that it holds
on subspaces of ${\cal S}_>$.  This ensures that vectors restricted to
subspaces also have positive length.  The subspaces of most relevance
to the Bethe-Salpeter equation are subspaces generated by equivalence classes
containing positive-time functions of one or two Euclidean space-time
variables.  Reflection positivity implies the following constraints  
on the two- and four-point Schwinger functions
\beq
\int d^4\x_1 d^4\x_2 
f^*_1 (\theta \x_1 ) 
S_{2} (\x_1,\x_{2} )
f_1 (\x_{2} ) \geq 0 . 
\label{eq:BI}
\eeq
\beq
\int d^4\x_1 \cdots d^4\x_{4} 
f^*_2 (\theta \x_2, \theta \x_1 ) 
S_{4} (\x_1,\x_2,\x_3,\x_4 )
f_2 (\x_{3}, \x_{4} ) \geq 0
\label{eq:BH}
\eeq

The relevant observation is that the two-
and four-point Schwinger functions must define a positive scalar
product on the subspaces defined above.  We test 
these conditions in section four.

To complete the construction of a relativistic quantum theory 
we need to construct a unitary representation of the Poincar\'e 
group on the physical Hilbert space.

The Poincar\'e group is a subgroup of the complex Poincar\'e group,
which also contains the real orthogonal group in four space-time
dimensions.  The infinitesimal generators of Euclidean transformations
and Poincar\'e transformations are related by complex multiplication.
Euclidean time-translations and rotations in Euclidean space-time
planes correspond to Poincar\'e time translations with imaginary
times, and rotationless Lorentz boosts with imaginary rapidity,
respectively.  While these finite transformations are unitary with
respect to a Euclidean scalar product, they are Hermitian with respect
to the physical scalar product (\ref{eq:BF}).  This is the reason that
the $\Theta$ appears in the definition of the physical scalar product.
The identity of the complex Lorentz group and complex $O(4)$ is
discussed in the appendix.

The infinitesimal forms of spatial translation, rotations,
Euclidean time translations, and Euclidean space-time rotations 
are used to identify the infinitesimal generators of the Poincar\'e
group.  The operators $H, \vec{P}, \vec{J}$ and $\vec{B}$  satisfy
the commutation relations of the Poincar\'e Lie algebra:
\[
\langle \x \vert H \vert f \rangle :=
\]
\beq
\{0 ,{\partial \over \partial
\x^0_{11}} f_1 (\x_{11}), 
\left ( {\partial \over \partial
\x^0_{21}} + {\partial \over \partial
\x^0_{22}} \right )  f_2 (\x_{21},\x_{22}), \cdots  \}
\label{eq:BJ}
\eeq
\[
\langle \x \vert \vec{P}  \vert f \rangle :=
\]
\beq
\{0 ,  - i {\partial \over
\partial \vec{\x_{11}}} f_1 (\x_{11}), 
 - i \left ( {\partial \over
\partial \vec{\x_{21}}}  + {\partial \over
\partial \vec{\x_{22}}} \right ) f_2 (\x_{21},\x_{22}), \cdots  \}
\label{eq:BK}
\eeq
\[
\langle \x \vert \vec{J}  \vert f \rangle :=
\]
\beq
\{0 , - i \vec{\x}_{11} \times
{\partial \over \partial \vec{\x}_{11}} f_1 (\x_{11}), 
  - i\left (  \vec{\x}_{21} \times
{\partial \over \partial \vec{\x}_{21}}  + \vec{\x}_{22} \times
{\partial \over \partial \vec{\x}_{22}} 
\right  ) f_2 (\x_{21},\x_{22}), \cdots  \}
\label{eq:BL}
\eeq
\[
\langle \x \vert \vec{B}  \vert f \rangle :=
\]
\[
\{0 , \left ( \vec{\x}_{11} 
{\partial \over \partial \x_{11}^0}- \x_{11}^0  {\partial \over
\partial \vec{\x}_{11}} \right )  f_1 (\x_{11}),
\]
\beq 
\left ( \vec{\x}_{21} 
{\partial \over \partial \x_{21}^0}- \x_{21}^0  {\partial \over
\partial \vec{\x}_{21}}+ \vec{\x}_{22} 
{\partial \over \partial \x_{22}^0}- \x_{22}^0  {\partial \over
\partial \vec{\x}_{22}} \right )  f_2 (\x_{21},\x_{22}), \cdots  \}.
\label{eq:BM}
\eeq
Elementary computations show that these operators are Hermitian with
respect to the physical scalar product (\ref{eq:BF}).  The Poincar\'e 
commutation
relations imply that these operators are the Hamiltonian, linear
momentum operators, the angular momentum operators, and the
rotationless Lorentz-boost generators.  For both $H$ and $\vec{B}$ the
Hermiticity follows because $\Theta$ changes the sign of the Euclidean
time.

When the model-Schwinger functions are covariant, rather than
invariant, and the discrete ``field'' indices transform with respect
to a finite-dimensional irreducible representation $D(A,B)$ of
$SL(2,C)\times SL(2,C)$ then the expression for the rotation and
Lorentz boost generators need to be modified
\beq
\left ( - i \vec{\x}_{11} \times
{\partial \over \partial \vec{\x}_{11}} \right ) \to 
\left ( - i \vec{\x}_{11} \times
{\partial \over \partial \vec{\x}_{11}} + \vec{\Sigma} \right )
\label{eq:BN}
\eeq
\beq
\left( \vec{\x}_{11} 
{\partial \over \partial \x_{11}^0}- \x_{11}^0  {\partial \over
\partial \vec{\x}_{11}} \right ) \to 
\left( \vec{\x}_{11} 
{\partial \over \partial \x_{11}^0}- \x_{11}^0  {\partial \over
\partial \vec{\x}_{11}}   +\vec{{\cal B}} \right ) 
\label{eq:BO}
\eeq
where 
\beq
\vec{\Sigma} = i \vec{\nabla}_{\phi} D(e^{{-i\over 2}\vec{\sigma}\cdot 
\vec{\phi}},  e^{{i\over 2}\vec{\sigma}^t\cdot 
\vec{\phi}})_{aa'}
\label{eq:BP}
\eeq
and 
\beq
\vec{{\cal B}} =  \vec{\nabla}_{\rho}D(e^{{-i\over 2}\vec{\sigma}\cdot 
\vec{\rho}},  e^{{-i\over 2}\vec{\sigma}^t\cdot 
\vec{\rho}})_{aa'} 
\label{eq:BQ}
\eeq
where the derivatives are evaluated at $\phi=\rho=0$ and 
$\vec{\sigma}^t$ is the transpose of $\vec{\sigma}$.

We have glossed over two technical points.  First, the Euclidean time
translations and Euclidean space-time rotations that were used to
construct $H$ and $\vec{B}$ do not map ${\cal S}_>$ to ${\cal S}_>$.
For the time translations this is addressed by considering only future
pointing Euclidean time translations, which do map ${\cal S}_>$ to
${\cal S}_>$.  Reflection positivity can be used to show that this
defines a contractive Hermitian semigroup.  Stone's theorem for
contractive Hermitian semigroups
\cite{riesz} guarantees that $H$ is a
self-adjoint operator on the physical Hilbert space.  The contractive
nature of the semigroup also implies that the Hamiltonian 
satisfies the spectral condition.  
For the Lorentz
transformations this is addressed by restricting the domain of the
transformations to successively larger positive-time convex cones,
${\Delta \x^0 \over \Delta \vert \vec{\x} \vert} < \tan (\phi)$, 
in ${\cal S}_>$ that map
into ${\cal S}_>$ for rotations through sufficiently small angles,
$\phi', \phi+\phi'< \pi/2$.  
Self-adjointness of the
generators on the physical Hilbert space is established by showing
that these restricted transformations are symmetric local semigroups
\cite{klein1}\cite{klein2}\cite{frohlich3}, which necessarily 
have self-adjoint generators.  

The second technicality is that it is necessary to establish that the
formal operators are well defined on the equivalence classes that
define Hilbert space vectors.  This is done by working on suitable
domains of functions and integrating by parts.  Specifically if $f \in
[0]_\sim$ and $X:{\cal S}_> \to {\cal S}_>$:
\beq
([g], \Theta S  [X f]) =
(g, \Theta S  X f)=
(Xg, \Theta S f)= 
([Xg], \Theta S [f])
= 0 
\label{eq:BR}
\eeq
which shows $f \in [0]_\sim$ implies $Xf \in [0]_\sim$, where $X$ can 
be any of the
operators, $\vec{P},\vec{J},\vec{B}$ or $H$. 

The result of this construction is a physical Hilbert space and a set
of ten self-adjoint operators that satisfy the Poincar\'e commutation
relations.  This shows how a collection of Euclidean-covariant,
reflection-positive model Schwinger functions can be used to directly
define a relativistic quantum dynamics satisfying the spectral
condition.  Note that all of the computations 
were performed using only Euclidean space-time variables; the corresponding 
model Wightman functions are not needed.

Particle exchange symmetry and microscopic locality put additional 
constraints on model Schwinger functions, but these additional 
constraints do not impact our stability analysis.

\section{One-Particle Systems:}

Since this Euclidean formulation of the relativistic quantum
theory is abstract and possibly unfamiliar, in this section we
show how it leads to familiar results for the case of a free
particle of mass $m$.

The two-point Schwinger function for a free field of mass $m$ 
is 
\beq
S_{2}(\x-\y) := 
{1 \over (2 \pi)^4} \int {d^4 \p \over \p^2 + m^2} e^{i \p \cdot (\x-\y)}.
\label{eq:CA}
\eeq

We first demonstrate that $S_{2} (\x-\y)$ is reflection positive
\cite{glimm}.
Let $f(\x)$ be any positive-time function of one Euclidean 
space-time variable.    
Reflection positivity requires (\ref{eq:BG}) 
\beq
(f,\Theta S_{2} f) := \int d^4\x d^4\y f (\x) S_{2}(\theta \x-\y) 
f(\y) \geq 0
\label{eq:CB}
\eeq
for all $f(\x)$ with positive Euclidean time support.

To demonstrate the inequality (\ref{eq:CB}) note  
\[
(f,\Theta S_{2}f) 
\]

\[ 
= {1 \over (2 \pi)^4} \int d^4\x d^4\y d^4\p f (\x) { e^{i \p \cdot (\theta 
\x-\y)}
\over \p^2 + m^2}  f(\y) 
\]
\beq
={1 \over (2 \pi)^4} \int d^4\x d^4\y d^4\p f (\x) { e^{-i \p_0 \cdot (\x_0+\y_0)
+ i \vec{\p} \cdot ( \vec{\x}-\vec{\y})}
\over  (\p^0 + i \omega_m (\vec{\p}\,))(\p^0 - i \omega_m (\vec{\p}\,))}
f(\y) 
\label{eq:CC}
\eeq
where
\beq
\omega_m (\vec{\p}) := \sqrt{ m^2 + \vec{\p} \cdot \vec{\p}}.
\label{eq:CD}
\eeq
The $\p_0$ integral is computed using the residue theorem.  The 
positivity of the 
Euclidean times means that the contour in the $\p_0$ integral should be closed
in the lower half $\p_0$-plane. 
The result of the contour integral is 
\beq
{1  \over 2} \int d^3 \p 
{\vert g(\vec{\p}) \vert^2 \over \omega_m (\vec{\p} \,) }
\geq 0
\label{eq:CE}
\eeq
where 
\beq
g(\vec{\p} ) := 
{1 \over (2\pi)^{3/2} } \int d^4y f(\y)
e^{-\omega_m (\vec{\p})y_0 -i \vec{\p} \cdot \vec{\y}}. 
\label{eq:CF}
\eeq
This is the standard Lorentz invariant scalar 
product for a particle of mass $m$, with momentum space wave function
$g(\vec{\p})$.  

The expressions for the Poincar\'e generators in 
(\ref{eq:BJ}-\ref{eq:BM}) 
act on the functions $f$, or more properly equivalences classes
$[f]_\sim$.   Using these covariant forms of the generators 
in equation (\ref{eq:CF}) leads to equivalent forms of the 
generators as operators acting on the wave functions $g(\vec{\p})$:
\beq
H = \omega_m (\vec{\p})
\label{eq:CG}
\eeq
\beq
\vec{P} = \vec{\p}.
\label{eq:CJ}
\eeq
\beq
\vec{J} = i \vec{\nabla}_\p \times \vec{\p}
\label{eq:CI}
\eeq
\beq
\vec{B} = i \omega_m \vec{\nabla}_\p .
\label{eq:CH}
\eeq
These are the familiar forms of the single particle 
Poincar\'e generators in the representation with the 
scalar product (\ref{eq:CE}).  

Similar results can be obtained for the case of spin $1/2$ particles.
For spin $1/2$ particles the Euclidean two-point 
Green function is
\beq
S_{2}(\x-\y) := 
{1 \over (2 \pi)^4} \int {d^4 \p} 
{m - p \cdot \gamma_e \over \p^2 + m^2} e^{i \p \cdot (\x-\y)}
\label{eq:CK}
\eeq
where 
\beq
i \gamma_{0e} = \gamma^0 = -\gamma_0; \qquad \gamma^i_e = \gamma^i .
\label{eq:CL}
\eeq 

In this case, because the Minkowski Green function is 
normally defined with a Dirac conjugate field rather than a 
Hilbert space adjoint, the $\gamma^0$ needs to be eliminated
from $S_2$ to get the continuation to the Wightman function 
that serves as the kernel of the Hilbert space scalar product.  
This can be achieved by introducing $\gamma^0$ as the spinor part of the 
$\Theta$ operator:
\[
(f,\Theta \gamma^0 S_{2}f) 
\]
\[ 
= {1 \over (2 \pi)^4} \int d^4\x d^4\y d^4\p f (\x) \,{ e^{i \p \cdot (\Theta 
\x-\y)}}
\gamma^0 {m - p \cdot \gamma_e \over \p^2 + m^2}
f(\y)  
\]
\beq
=\int d^3 \p g^{\dagger}(\vec{\p}) {\Lambda_+ (p) \over (2 \pi)^3} 
g(\vec{\p})  
\label{eq:CM}
\eeq
where
\beq
\Lambda_+ (p) := {\omega_m (\vec{p}) +  
\gamma^0 \vec{\gamma} \cdot \vec{p} - m \gamma^0 
\over 2 \omega_m (\vec{p})} 
\label{eq:CN}
\eeq
is the positive energy Dirac projector and
\beq
g(\vec{p}) := \int d^4\x \, 
e^{- \omega_m (\vec{p}) \x_0 
- i \vec{\p} \cdot \vec{\x}} f(\x) 
\label{eq:CO}
\eeq
as before, except in this case $f(\x)$ is a four-component 
covariant wave function.  Note that 
\beq
\Lambda_+ (p) = {\chi (p) \chi^{\dagger} (p)
\over 2 \omega_m (\vec{p}) }
\label{eq:CP}
\eeq
is a Hermitian matrix with  
\beq
\chi (p) = 
\sqrt{m}\gamma^0 u_m(\vec{\p})
\label{eq:CQ}
\eeq
where $u_m (\vec{\p})$ a Dirac $u$-spinor.  The function $\chi (p)$
is a $2 \times 4$ matrix that satisfies the intertwining relation
\beq
\Lambda \chi (p) = \chi (\Lambda p) R_w (\Lambda ,p)  
\label{eq:CR}
\eeq
where $R_w (\Lambda ,p)$ is a Wigner rotation.  This matrix
intertwines Dirac spinor representations of the Lorentz group with
positive-mass positive-energy irreducible representation of the
Poincare group.

If we define 
\beq
\psi (p) := {1 \over (2 \pi )^{3/2}}\chi^{\dagger}(p) g(\vec{p})  
\label{eq:CS}
\eeq
the scalar product
\beq
\langle \psi \vert \psi \rangle := 
\int {d^3p \over 2 \omega_m (\vec{p})}  
\psi^{\dagger} (\vec{p}) \psi (\vec{p})
\label{eq:CT}
\eeq
has the standard form of a mass $m$, spin $1/2$ irreducible representation 
of the Poincar\'e group.  Note that two component spinors characteristic
of spin 1/2 positive mass irreducible representations of the Poincar\'e group 
arise from the factorization (\ref{eq:CP}). 

The intertwining relations (\ref{eq:CR}) imply that when the generators 
(\ref{eq:BJ}-\ref{eq:BM}) are transformed to act on the 
Poincar\'e irreducible 
wave functions $\psi (p)$ that the 
generators take on the standard forms 
\beq
\vec{B} = i \omega_m \vec{\nabla}_\p + 
{1 \over m + \omega_m (p) } \vec{p} \times \vec{j} 
\label{eq:CU}
\eeq
\beq
\vec{J} = i \vec{\nabla}_\p \times \vec{\p} + \vec{j}
\label{eq:CV}
\eeq
where $\vec{j}$ is the canonical spin operator that acts on the 
two-dimensional range of $\chi$
\beq
\vec{\Sigma} \chi (p) = \chi (p) \vec{j}_c 
\label{eq:CW}
\eeq
where $\vec{\Sigma}$ is defined by (\ref{eq:BP}).

This shows how the abstract Euclidean formulation of the relativistic quantum
dynamics associated with a set of model Schwinger functions given in section
three leads to the standard Minkowski description of the dynamics of a
single particle in terms of irreducible representations of the 
Poincar\'e group.

Since the many-point Schwinger functions for a system of free
particles are made up out of sums of tensor products of two-point
functions, this result implies reflection positivity for the full set
of free-particle Schwinger functions.

\section{Stability}  

In this section we investigate the stability of reflection positivity.
This is the main result in this paper.  Our analysis is limited to the
necessary conditions for reflection positivity given by equations
(\ref{eq:BH}) and (\ref{eq:BI}).

To gain some insight into the problem we first consider a toy $2\times 2$
matrix model.  We use this model to check stability of reflection 
positivity in the simplest possible case.

The problem is to consider a model of the inhomogeneous 
Euclidean  Bethe-Salpeter equation 
\beq
S= S_0 + S_0 K S
\label{eq:DA}
\eeq
where $S_0$ is positive and reflection positive.  Note that while the
positivity of $S$ is not required, for Schwinger functions of scalar
fields it is used to derive bounds needed to prove the spectral condition
\cite{glimm}.  What are the restrictions on $K$ that preserve these
properties?

To motivate the matrix model consider the
quadratic form $(f,\Theta S f)$ and model $f$ by a constant $f_c$
times $\delta (\vec{\x})\delta (\x^0 -\t)$ with $\t$ positive.  We also
consider the time reflected point $\t'=\theta \t=  - \t$.  We treat $f$
as a column vector with the upper component corresponding to the value
of $f=f_c$ at Euclidean time $\t$ and the lower component corresponding 
to the value $f=0$ at the Euclidean time-reversed point $\theta \t = -\t$.

In this model we define
\beq
\Pi_> := 
\left (
\begin{array}{cc}
I & 0 \\
0 & 0 
\end{array} 
\right )
\label{eq:DB}
\eeq
and
\beq
\Theta :=
\left (
\begin{array}{cc}
0 & I \\
I & 0 
\end{array} 
\right ) .
\label{eq:DC}
\eeq
The quadratic form becomes
\beq
(f,\Theta S f) = 
(f_c,0)
\left (
\begin{array}{cc}
0 & I \\
I & 0 
\end{array} 
\right )
\left (
\begin{array}{cc}
s(\vec{0}, 0) & s(\vec{0}, 2\t) \\
s(\vec{0},-2\t) & s(\vec{0}, 0)
\end{array} 
\right ) 
\left (
\begin{array}{c}
f_c \\
0 
\end{array} 
\right ) = f_c^2 s(\vec{0}, -2\t )
\label{eq:DE}
\eeq
where $(f_c,0)$ represents a row vector.  In this case reflection
positivity on this one-dimensional subspace, analogous to
(\ref{eq:BH}), requires that $s(\vec{0},-2\t) >0$.

We write this in a more abstract form by defining
\beq
s_0 := \left (
\begin{array}{cc}
s_{011} & s_{012} \\
s_{021} & s_{022}
\end{array} 
\right ) = \left (
\begin{array}{cc}
s(\vec{0},0) & s(\vec{0},2\t) \\
s(\vec{0},-2\t) & s(\vec{0},0)
\end{array} 
\right ). 
\label{eq:DF}
\eeq
In this notation reflection positivity means
\beq
\Pi_> \Theta S_0 \Pi_> =
\left (
\begin{array}{cc}
s_{021} & 0 \\
0  & 0 
\end{array} 
\right ) \geq 0
\label{eq:DG}
\eeq
or $s_{021}>0$. Euclidean invariance requires
\beq
s_0 = \Theta s_0 \Theta 
\label{eq:DH}
\eeq
which implies that $s_{0ij}$ are real and satisfy
$s_{011}=s_{022}$ and $s_{012}=s_{021}$.
It is straightforward to show  
the requirements of positivity and reflection positivity 
in this model are satisfied if 
\beq
s_{011} > s_{012} >0  
\label{eq:DI}
\eeq
which means that the matrix $S_0$ must have positive elements with the
diagonal ones being larger than the off-diagonal ones.  This
condition must hold for any $2\times2$ sub-matrix associated with times
$\pm \t$.  For larger matrices it is only a necessary condition;
however this suggest that in the general case there is a growth
condition limiting the size of off diagonal elements relative to
diagonal elements.

The next step is to add a perturbation using a kernel $K$.  In
this case we model the Bethe-Salpeter kernel $K$ by an ``Euclidean'' 
invariant $2 \times 2$ Hermitian matrix.    If 
\beq
K=
\left (
\begin{array}{cc}
k_{11} & k_{21} \\
k_{12} & k_{22}
\end{array} 
\right ) 
\label{eq:DJ}
\eeq
Hermiticity and Euclidean invariance require 
that $K$ is real, and $k_{11}=k_{22}$ and 
$k_{12}=k_{21}$.  This means that ``kernels'' $K$ can be parameterized 
by vectors in the two-dimensional $(k_{11}, k_{12})$ plane.
The equation for $S$ is 
\beq
\left (
\begin{array}{cc}
s_{11} & s_{21} \\
s_{12} & s_{22}
\end{array} 
\right ) =
\left (
\begin{array}{cc}
s_{011} & s_{012} \\
s_{012} & s_{011}
\end{array} 
\right ) +
\left (
\begin{array}{cc}
s_{011} & s_{012} \\
s_{012} & s_{011}
\end{array} 
\right ) 
\left (
\begin{array}{cc}
k_{11} & k_{12} \\
k_{12} & k_{11}
\end{array} 
\right ) 
\left (
\begin{array}{cc}
s_{11} & s_{21} \\ 
s_{12} & s_{22}
\end{array} 
\right ) .
\label{eq:DK}
\eeq
This equation can be solved for $S$  and the conditions for positivity and 
reflection positivity are found to be:
\beq
k_{11} + k_{12} < 
{1 \over s_{011} + s_{012}}
\label{eq:DL}
\eeq
\beq
k_{11} <  {s_{011}
\over \mbox{det}(S_0)}=    {s_{011}
\over s_{011}^2 - s_{012}^2}    
\label{eq:DM}
\eeq
\beq
k_{12} > - {s_{012} \over \mbox{det}(S_0)}=-
{s_{012}
\over s_{011}^2 - s_{012}^2} .   
\label{eq:DN}
\eeq
The important property of these inequalities is that they define a
region that contains an {\it open set containing the origin} in the
$(k_{11}, k_{12})$ plane.  This means that reflection positivity is
preserved in this model for sufficiently small Bethe-Salpeter kernels.
In this trivial model the reflection positivity condition is stable 
with respect to small variations about the unperturbed system.  
This simple construction provides clues about the key elements of a 
general stability construction.

The next step is to consider the actual 
inhomogeneous Euclidean Bethe-Salpeter
equation. Abstractly we still have the operators $\Pi_>$, $\Theta$,
$S_0$, $S$, and $K$.  Euclidean covariance requires
\beq
[\Theta ,S] = [\Theta ,K] = [\Theta , S_0] =0 .
\label{eq:DO}
\eeq
Under the conditions that $S_0$ has an inverse  
the operator $T$ is defined by 
\beq
T= S_0^{-1} + S_0^{-1} S S_0^{-1}.
\eeq 
This can be used to write $S$ in the
solved form
\beq
S = S_0 + S_0 T S_0 .
\label{eq:DP}
\eeq
It follows from (\ref{eq:AAK}) and (\ref{eq:DP}) that 
the Bethe-Salpeter $T$-operator can be determined directly  
by solving the equation 
\beq
T = K + K S_0 T .
\label{eq:DQ}
\eeq
If the K\"all\'en-Lehmann representation of the two-point function has a 
mass gap and the Euclidean norm of $K$ is sufficiently small  
this equation has a unique solution.

If $f \in {\cal S}_>$ is a function of two Euclidean space-time 
variables then the (norm)$^2$ of the vector $[f]_\sim$ in the {\it interacting}
theory is given by
\beq
\Vert [f]_\sim \Vert^2 =
( \Theta \Pi_> f ,   (S_0 + S_0 T S_0) 
\Pi_> f) . 
\label{eq:DR}
\eeq
This can be written in the form
\beq
\Vert [f]_\sim \Vert^2 = ( \Pi_> f ,   (\Theta S_0 + S_0 \Theta T S_0) 
\Pi_> f)  
\label{eq:DS}
\eeq
where we have used the Euclidean invariance of $S_0$.  Equation
(\ref{eq:BH}) {\it requires} that this form is {\it non-negative} if
it is interpreted as the norm of a vector in an underlying quantum
theory.  The main results of the paper are contained in
three theorems.

\noindent {\bf Theorem 1:} If $f\not=0$ satisfies
\beq
( \Pi_> f ,   \Theta S_0  
\Pi_> f) = 0 
\label{eq:DT}
\eeq
then there are arbitrarily small Euclidean covariant Bethe-Salpeter 
kernels that violate reflection positivity of $S$. 

To prove Theorem 1 note that (\ref{eq:DT}) implies 
that the surviving contribution to equation (\ref{eq:DR}) is 
\beq
\Vert [f]_\sim \Vert^2 = ( \Pi_> f , S 
\Pi_> f)  = 
( S_0 \Pi_> f , ( \Theta T)  S_0  \Pi_> f).
\label{eq:DU}
\eeq
Let
\beq 
\chi = S_0 \Pi_> f
\label{eq:DV}
\eeq
which gives
\beq
\Vert [f]_\sim \Vert^2 = ( \Pi_> f , \Theta S 
\Pi_> f)  = (\chi , \Theta T \chi ) . 
\label{eq:DW}
\eeq
Note that $\chi$ is not an element of ${\cal S}_>$.
Since the free $S_0$ is invertible on the full Euclidean space, $\chi$
is not zero.  An arbitrarily small $\Theta T$ with Euclidean covariant
$T$ can be chosen to have non-zero matrix elements on the subspace
spanned by the $\chi$'s.  This quantity must be non-negative for this
to represent the square of the Hilbert-space norm in the
interacting theory.

We are concerned with the stability of the sign of (\ref{eq:DW})
for sufficiently small $T$.  If this expression is negative we get a
violation of the necessary condition (\ref{eq:BH}) for reflection
positivity; if it is positive, then $T \to -T$ gives a negative
result, independent of the size of the perturbation $T$. Note that $T
\to -T$ is equivalent to $K \to K_-$ where $K_-$ is the solution to
the integral equation
\beq
K_- = -K+ 2 K S_0 K_- .
\label{eq:DX}
\eeq
This will be small if $K$, considered as an operator, has a sufficiently 
small Euclidean norm.  This completes the proof of
Theorem 1.

This is the stability problem that is the key cause for concern in
this paper.  The reason that the problem did not appear in the $2\times2$
matrix model is because the ``unperturbed model'' was implicitly
constructed to have no null space.

The following theorems show that the condition for the instability 
to occur is satisfied if $S_0$ is constructed out of free 
two-point Schwinger functions.  They also show that the exact two-point
Schwinger functions do not have this property. 

\bigskip
\noindent{\bf Theorem 2:}  The null equivalence class of $S_0$ , $[0]_\sim$,
contains an infinite number of functions if $S_0$ is the Schwinger function
of a {\it free} field theory: 

\bigskip
\noindent{\bf Theorem 3:}  The null equivalence class of $S_0$ , $[0]_\sim$,
contains no non-zero functions if $S_0$ has a K\"all\'en-Lehmann 
weight with an
absolutely continuous component of its mass spectrum.

\bigskip
To prove these theorems we note that 
the general form of the two-point Schwinger function for a 
scalar field theory is given by its K\"all\'en-Lehmann representation 
\beq
S_0 (\x-\x') =
{1 \over (2 \pi)^2 } \int d^4\p dm
\rho (m) { e^{i \p^0 (\t-\t') + 
i \vec{\p}\cdot (\vec{\x}-\vec{\x}') }
\over  \p^2 + m^2 }.
\label{eq:DY}
\eeq
A function $f(\vec{\x},\t)$ with positive-time support is in the null 
equivalence class of 
$S_0$ on the range of $\Pi_>$ if and only if
\beq
(f,\Theta S_0 f) = 0 .
\label{eq:DZ}
\eeq
  
To see what this means note that 
\[
(f,\Theta S_0 f) = 
\]
\beq
\int f^* (\vec{\x},-\t){1 \over (2 \pi)^2 } \int d^4\p dm
\rho (m)   \p {e^{i \p^0 (\t-\t') - i 
\vec{\p}\cdot (\vec{\x}-\vec{\x}') }
\over  (\p^0 - i \omega_m (\vec{\p}\,))
(\p^0 + i \omega_m (\vec{\p}\,))}
f(\vec{\x}',\t') d^4\x d^4\x' .
\label{eq:DAA}
\eeq
Direct calculation of this gives
\beq
(f,\Theta S_0 f) =  \int d^3\p dm \rho (m) \left \vert \int d\t   
\tilde{f} (\vec{\p},\t) {2 \pi e^{- \omega_m (\vec{\p})\t}
\over \sqrt{\omega_m (\vec{\p})}} \right \vert^2
\label{eq:DAB}
\eeq
where 
\beq
\tilde{f} (\vec{\p},\t) := {1 \over (2 \pi)^{3/2} }
\int d^3 \x e^{i \vec{\p}\cdot \vec{\x} } f(\vec{\x},\t)  .
\label{eq:DAC}
\eeq
This will vanish if and only if 
\beq
\int
d\t \tilde{f} (\vec{\p},\t) {2 \pi e^{- \omega_m (\vec{\p})\t}
\over \sqrt{\omega_m (\vec{\p})}} =0
\label{eq:DAD}
\eeq
for all values of $\vec{\p}$ and all $m$ in the spectrum of the
K\"all\'en-Lehmann weight, $\rho (m)$, of the two-point function.  
A necessary and
sufficient condition for $f$ to represent a function in the 
null equivalence class of $S_0$ is
\beq
I=\int
d\t \tilde{f} (\vec{\p},\t) e^{- \omega_m (\vec{\p})\t} = 0
\label{eq:DAF}
\eeq
for all $\vec{\p}$ and all $m$ in the spectrum of the 
K\"all\'en-Lehmann weight. 

We first consider the free field case where there is only a single 
mass ($\rho (m) =\delta (m-m_0)$).  We show how to 
construct a large class of functions
$\tilde{f}(\vec{\p},\t)$ with support on compact 
positive Euclidean time intervals $[a,b]$, $0<a <b<\infty$,
in the null equivalence class of $S_0$.
Let $\chi (\t)$
be a Schwartz function with support on $[a,b]$ satisfying the 
normalization condition
\beq
\int_a^b d\t \chi (\t) = 1,
\label{eq:DAG}
\eeq
let $\tilde{g} (\vec{\p}) e^{b \omega_m (\p)}$ be a Schwartz 
function of $\vec{\p}$
and let
\beq
h(\vec{\p}) := \int d\t e^{- \omega_m (\vec{\p})\t }\chi (\t)  .
\label{eq:DAH}
\eeq
Define
\beq
\tilde{f}(\vec{\p},\t):= \chi (\t) \tilde{g} (\vec{\p}) \times
[1  -e^{\omega_m (\p)t}h(\vec{\p})]. 
\label{eq:DAI}
\eeq
By construction $\tilde{f}(\vec{\p},\t)$ is a non-trivial 
function with compact positive-time support on $[a,b]$  
satisfying $(f, \Theta S_0 f) =0$.

In the Euclidean Bethe-Salpeter equation $S_0$ is a sum of products of
two-point Schwinger functions.  From the discussion above, if the
individual Schwinger functions have single masses then it is possible to
find functions with support in any compact positive time interval that
are in the null equivalence class of $S_0$.  Choosing products of functions 
with disjoint positive time support it is possible to find
functions in the range of $\Pi_>$ that are in the null equivalence class 
of the tensor product of two $\Theta S_0$'s.    

On the other hand it is clear that $\Theta S_0 f$ is non-zero 
since for $g= \Theta f$ 
\[
(g, \Theta S_0 f) = (f,S_0 f) = {1 \over (2 \pi)^4} \int d^4 \p
d^4 \x d^4\y
f^* ( \vec{\x},\t_x) {e^{i \p \cdot (\x-\y)} \over \p^2 + m^2} 
f ( \vec{\y},\t_y)=
\]
\beq 
\int d^4\p {\vert \hat{f} (\vec{\p} ,\p^0) \vert^2  \over \p^2 + m^2 
} > 0,
\label{eq:DAJ}
\eeq
where $\hat{f} (\vec{\p} ,\p^0)$ is the four dimensional Fourier 
transform of $f( \vec{\x},\t_x)$.  This implies that
for any $f$ in the null equivalence class 
of $S_0$ there are functions $g$ with no support restrictions 
that satisfy 
\beq
(g, \Theta S_0 f) \not= 0.
\label{eq:DAK}
\eeq
If there are exchange contributions to the Green functions,
we have 
\[
([f_1]_\sim[f_2]_\sim , \Theta S_0 [f_1]_\sim[f_2]_\sim )=
\]
\beq
([f_1]_\sim , \Theta S_{01} [f_1]_\sim )
([f_2]_\sim , \Theta S_{01} [f_2]_\sim )+
([f_1]_\sim , \Theta S_{01} [f_2]_\sim )
([f_2]_\sim , \Theta S_{01} [f_1]_\sim )
\label{eq:DAL}
\eeq
which is zero if $[f_1]_\sim=[f_2]_\sim=[0]_\sim$.

It follows that if $S_0$ is the $S_0$ of a free field that $
\Pi_> \Theta S_0 \Pi_>$
has a non-trivial null space on the range of $\Pi_>$.
This completes the proof of Theorem 2.

The analysis above also applies to the spin $1/2$ case because
the $\t$ dependence in (\ref{eq:DAF}) and (\ref{eq:CO}) is 
identical.   

This shows that solutions of the Euclidean Bethe-Salpeter equation
formulated with a free $S_0$ are {\it not} reflection-positivity
stable with respect to small perturbations.  This means one can alway
find arbitrarily small Bethe-Salpeter kernels that make 
$\Vert [f]_\sim  \Vert^2 = ([f]_\sim , \Theta
S [f]_\sim )<0$, which violates reflection positivity.

Since the two-point functions that appear in the {\it exact} Bethe-Salpeter 
equation of a local field theory are not generally sums of products of 
free Schwinger functions, it is worth investigating if these more 
realistic two-point functions 
have a non-trivial null equivalences classes.  
Theorem 3 addresses this question. 

To prove Theorem 3 fix $\vec{\p}$.
The condition for $\tilde{f} (\vec{\p},\t)$ to be in the null equivalence 
class of $S_0$ is (\ref{eq:DAF}).

If $\tilde{f} (\vec{\p},\t)$
is a Schwartz function in $\t$ with support for positive $\t$ 
then for fixed $\vec{\p}$  
\beq
F(z ) := \int d\t \tilde{f} (\vec{\p},\t) e^{- z \t}
\label{eq:DAM}
\eeq
is an analytic function for $\Re (z) >0 $.  As $m$ varies 
continuously in the spectrum of the K\"all\'en-Lehmann weight 
$z=\omega_m (\vec{p}) =z(m) $ traces out a real interval in the 
domain of analyticity of $F(z)$, where $F(z(m))=0$.  Since $F(z)$
is analytic in the right half plane, it must be identically zero
on the entire domain of analyticity.  If $\tilde{f} (\vec{\p},\t)$
is a Schwartz function in $\t$ for fixed $\vec{\p}$ this is continuous 
on the boundary as $y \to 0$.  It follows that 
\beq
\int d\t \tilde{f} (\vec{\p},\t) e^{-iyt} =
\lim_{x \to 0^+} F(x+iy  ) = 0
\eeq
which means that the Fourier transform of $\tilde{f} (\vec{\p},\t)$ in $\t$
is identically zero.  This same argument can be applied to each
$\vec{\p}$.  This proof can be extended to the case that $\tilde{f}
(\vec{\p},\t)$ is a tempered distribution in $\t$ \cite{simon3}. 

Since this can be done for any $\vec{\p}$ it follows that there
are no non-zero functions in the null equivalence class of $S_0$
if the spectrum of K\"all\'en-Lehmann weight has any absolutely
continuous component.  This completes the proof of Theorem 3.

This is an encouraging result that leaves open the possibility that a
stability result might be possible if the Bethe-Salpeter equation is
formulated with a more realistic two-point function in the driving
term.

\section{Conclusion} 

In this paper we proved that the solution of the Euclidean Bethe-Salpeter 
equation 
\beq
S= S_0 + S_0 K S 
\label{eq:EA}
\eeq
with $S_0$ a {\it free field} Schwinger function can violate
reflection positivity for arbitrarily small Euclidean covariant
kernels $K$.  When reflection positivity is violated, the standard
axiomatic construction of the physical Hilbert space leads to a
pathological inner product with negative (norm)$^2$ vectors, making it
impossible to give the theory a quantum mechanical interpretation.

The basis of the instability is simple to understand.  In quantum
field theory the quantum mechanical scalar product can be expressed
directly in terms of the Schwinger functions as
\beq
\langle f \vert g \rangle := (\Pi_> f , \Theta S \Pi_> g) .
\label{eq:EB}
\eeq
The  Schwinger function can be expressed using a cluster expansion
as the sum of a linked term and unlinked term
\beq
S= S_0 + S_0 T S_0 
\label{eq:EC}
\eeq
where $T$ is Euclidean covariant.  The Euclidean Bethe-Salpeter equation 
generates the linked terms in terms of the Bethe-Salpeter kernel, $K$.
If a test function $f$ in the range of $\Pi_>$ satisfies
$(\Pi_> f , \Theta S_0 \Pi_> f)=0$ then the expression for the 
square of the norm of the corresponding vector is 
\beq
\Vert f \Vert^2 := (\Pi_> f , S_0 \Theta T S_0 \Pi_> f) =
(S_0 \Pi_> f , \Theta T S_0 \Pi_> f). 
\label{eq:ED}
\eeq
If we pick an arbitrary Euclidean covariant $T$ satisfying $\Vert T
\Vert_e < \epsilon$ that makes the above expression non-zero, then
either the (norm)$^2$ is negative or it can be made negative by changing
the sign of $T$.  In either case we end up with an instability for 
an arbitrarily small perturbation $T$.  Similar remarks apply 
to the kernel $K$. 

The existence of the instability depends on the existence of functions
of Euclidean space-time variables with Euclidean-time support in the
range of $\Pi_>$ that satisfy $(f,\Theta S_0 f)=0$.  We exhibited a large
class of these functions for the case that $S_0$ is the $S_0$ of a
{\it free} field theory.  Conversely, we argued that there are no such
functions when the K\"all\'en-Lehmann 
weight of the two-point function includes a
continuous mass spectrum.  This suggests that the instability only
impacts model Bethe-Salpeter equations where the driving term is
replaced by the free $S_0$

The problems with the sign of the Hilbert space norm are unrelated to
sign problems that sometimes occur with the normalization of
Bethe-Salpeter amplitudes \cite{nakanishi}.  The signs associated with the
Bethe-Salpeter normalization condition are directly related to the
normalization of the Green function, however they say nothing about
the underlying Hilbert space of the theory.

Negative norms can appear in gauge theories.  In the case of gauge
theories negative norms arise because the space generated by applying
polynomials of the field operators to the vacuum includes unphysical
degrees of freedom.  The problem identified in this paper occurs in
models of theories where polynomials in the fields applied to the
vacuum only generate physical states.  Thus the negative norms
identified in this paper are associated with vectors that should
represent physical states.  An investigation of the analogous
stability question in a gauge theory is beyond the scope of this
paper.

Another question is whether the constraint of Euclidean covariance is
the appropriate minimal constraint on the model Bethe-Salpeter
kernels. We were unable to identify other generic constraints which are
motivated by the structure of local quantum field that would prevent the
instability.
    
Our interest in this instability arose from attempts to construct a
robust class of relativistic quantum models based on solutions of a
Euclidean Bethe-Salpeter equation with the driving term being the
$S_0$ of a free field theory.  One goal was to identify a class of
model Euclidean-Bethe Salpeter kernels, for example small,
Euclidean-covariant separable kernels, that could be used to construct
relativistic quantum models, under the assumption that the resulting
model model-Schwinger functions have the same relation to the
underlying quantum theory as the exact Schwinger functions.  The
identification of the instability in this paper shows that this
problem has no solution if the kernels are allowed to be arbitrary
Euclidean covariant kernels with sufficiently small Euclidean norms.

This work suggests that an interesting problem is to investigate the
more realistic case, where the weight of the K\"all\'en-Lehmann
representation of the two-point function has a non-empty absolutely
continuous spectrum.  The absence of a null space suggest that it
might be possible to generalize the analysis of the matrix model to to
find bounds on the kernel, expressed in terms of $S_0$, that might
lead to stability.  A result of this type would provide useful
restrictions on model Bethe-Salpeter kernels that ensure a
relativistic quantum interpretation.  

The authors have benefited from useful discussions
with P.E.T. Jorgensen, F. Coester, M. Fuda and D. Phillips.

\vfill\eject

\bigskip

\appendix{Appendix: The Complex Euclidean/Lorentz Group}

\bigskip 

The complex Lorentz group \cite{wightman2} and complex $O(4)$ are 
the same group.  The covering group of both groups is 
$SL(2,C) \times SL(2,C)$.
To illustrate this connection let $x$ be a real Lorentz four vector.  Let
\beq
\sigma_{\mu} = ( I, \sigma_1 , \sigma_2 , \sigma_3) 
\eeq
be the $2 \times 2$ identity matrix and the three $2\times 2$ Pauli
spin matrices.  Define the Hermitian $2\times 2$ matrix
\beq
X := x^{\mu} \sigma_{\mu} = 
\left (
\begin{array}{cc}
x^0 + x^3 & x^1 - i x^2 \\
x^1 + i x^2 & x^0 - x^3 
\end{array}
\right ).
\eeq
This definition implies:
\beq
x^{\mu} = {1 \over 2}\mbox{Tr} [\sigma^{\dagger}_{\mu} X] .
\label{eq:appa}
\eeq
The determinant of the matrix $X$, 
\beq
\det (X) = -x\cdot x = (x^0)^2 - \vec{x} \cdot \vec{x},
\eeq
is the Lorentz invariant length of $x$.  Also note that 
for real $x$, $X^{\dagger} = X$.

The set of non-trivial linear transformations 
that preserve both $\det (X)$ and $X=X^{\dagger}$ have the form:
\beq 
X' = \Lambda X \Lambda^{\dagger}
\label{eq:appb}
\eeq
where $\Lambda$ is any complex $2 \times 2$ matrix with $\det (\Lambda) =1$.
These matrices have a $2$ to $1$ correspondence with the real Lorentz 
transformations continuously connected to the identity:
\beq
\Lambda^{\mu}{}_{\nu}  := 
{1 \over 2} \mbox{Tr} [\sigma_{\mu} \Lambda \sigma_{\nu} 
\Lambda^{\dagger}]
\eeq
with $\Lambda$ and $-\Lambda$ corresponding to the same 
real Lorentz transformation.  This can be derived by multiplying
(\ref{eq:appb}) by $\sigma_{\mu}$ and taking 
the trace using the trace formula (\ref{eq:appa}). 
Rotations correspond to the case that $\Lambda$ is unitary while rotationless
boosts correspond to the case the $\Lambda$ is a positive matrix.

Including indices on $X \to X_{ab}$ the transformation properties are
\beq
X'_{ab} = \Lambda_{a}{}^{a'} \otimes (\Lambda^*)_{b}{}^{b'}X_{a'b'}.
\eeq

This shows that a four vector transforms covariantly with respect to
$\Lambda \otimes \Lambda^*$.  For a general complex $\Lambda$, $\Lambda$ 
and $\Lambda^*$ define
inequivalent representations of $SL(2,C)$; they cannot be related by a
similarity transformation.  These are the fundamental representation
of $SL(2,C)$ and are the building blocks of all (finite dimensional)
spinor and tensor
representations of the Lorentz group.

In a similar manner let $\x$ be a real Euclidean four vector.  Let 
\beq
\sigma_{e\mu} = ( iI, \sigma_1 , \sigma_2 , \sigma_3)
\eeq
and
\beq
{\sf X} := \x^{\mu} \sigma_{e\mu} = 
\left (
\begin{array}{cc}
i\x^0 + \x^3 & \x^1 - i \x^2 \\
\x^1 + i \x^2 & i\x^0 - \x^3 
\end{array}
\right ).
\eeq
This can be inverted using multiplication by the Pauli 
matrices and taking traces:
\beq
\x^{\mu} = {1 \over 2} \mbox{Tr} (\sigma^{\dagger}_{e\mu} {\sf X} ).
\label{eq:appc}
\eeq

Note that 
\beq
\det ({\sf X}  ) = -\x\cdot \x = - (\x^0)^2 - \vec{\x} \cdot \vec{\x}
\eeq
which is $(-)$ the Euclidean invariant $($length$)^2$ of $\x$.  Also note 
for real $\x$ that ${\sf X} = - \sigma_2 {\sf X}^{*} \sigma_2$.

This condition is preserved for 
\beq
{\sf X}' = A {\sf X} B^{t} 
\label{eq:appd}
\eeq
provided both $A$ and $B$ are unitary.  In terms of the individual 
Euclidean components
\beq
E^{\mu}{}_{\nu} = E(A,B)^{\mu}{}_{\nu} :=
{1 \over 2} \mbox{Tr} [\sigma_{e\mu}^{\dagger} A \sigma_{e\nu} B^t].
\eeq
This can be derived by multiplying
(\ref{eq:appd}) by $\sigma_{e\mu}^{\dagger}$ and taking 
the trace using the trace formula (\ref{eq:appc}).

Including indices on the ${\sf X} \to {\sf X}_{a b}$ then 
the transformation properties are
\beq
{\sf X}_{a  b}  \to {\sf X}'_{a  b} =
A_a{}^{a'}\otimes B_{ b}{}^{ b'}{\sf X}_{a' b'}.
\eeq
This shows that a Euclidean four vector transforms covariantly with 
respect to $A\times B$.

The restrictions $X'= AXA^{\dagger}$ with $\det (A)=1$ and 
$\X'= A\X B^t$ with $A$ and $B$ unitary are designed to keep the components 
of the Minkowski or Euclidean four vectors real.  For $A$ and $B$
arbitrary complex $2\times 2$ matrices with $\det (A)= \det (B) =1$
the invariant length (Minkowski or Euclidean) is still preserved.
What changes is that the individual components of the vectors are 
complex.  The complex transformations that act on Lorentz or Euclidean 
four vectors are
\beq
\Lambda (A,B)^{\mu}{}_{\nu} :=
{1 \over 2} \mbox{Tr} [\sigma_{\mu} A \sigma_{\nu} B^t]
\eeq
and
\beq
E(A,B)^{\mu}{}_{\nu} :=
{1 \over 2} \mbox{Tr} [\sigma_{e\mu}^{\dagger} A \sigma_{e\nu} B^t] .
\eeq
These two representation differ by a similarity transformation that multiplies 
Minkowski time by $i$ to get the Euclidean time.  This means the any four 
vector has Euclidean and Minkowski components, $\x^{\mu}$ and $x^{\mu}$
respectively.  Under $SL(2,C) \times SL(2,C)$ they transform like
\beq
x^{\mu} \to x^{\prime \mu} = \Lambda (A,B)^{\mu}{}_{\nu} x^{\nu}
\eeq
or equivalently
\beq
\x^{\mu} \to \x^{\prime \mu} = E(A,B)^{\mu}{}_{\nu} \x^{\nu}.
\eeq
The Lie algebra of the Lorentz group can be derived by
considering the appropriate infinitesimal complex Euclidean transformations

Transformation properties of Euclidean Green functions can be determined by the
transformation properties of the corresponding Minkowski functions by replacing
the fundamental representations $(\Lambda, \Lambda^{*})$ by the pair of 
$SU(2)$ matrices $(A,B)$.

In applications it is important to understand the transformation
properties of tensor and spinor quantities with respect to Euclidean
transformations, given a knowledge of the transformation properties of
the corresponding Minkowski quantities.  In general the transition is
made from the finite dimensional representation $D(\Lambda ,
\Lambda^*)$ of the Lorentz group to the finite dimensional
representation $D(A,B)$ of the four dimensional orthogonal group by
making the replacements $\Lambda \to A$ and $\Lambda^* \to B$.

\end{document}